%% file: main.tex
\documentclass{IOS-Book-Article}     
\begin{document}
\begin{frontmatter}          
%
\title{LLM-Augmented Agent-Based Modelling for Social Simulations: Challenges and Opportunities}

\author{\fnms{Önder} \snm{Gürcan}}
\address{
    Norwegian Research Center AS (NORCE),\\
    Universitetsveien 19, Kristiansand, Norway\\
    \email{ongu@norceresearch.no}
}
\runningauthor{}
%
%
\begin{abstract}
As large language models (LLMs) continue to make significant strides, their better integration into agent-based simulations offers a transformational potential for understanding complex social systems. 
However, such integration is not trivial and poses numerous challenges.
Based on this observation, in this paper, we explore architectures and methods to systematically develop LLM-augmented social simulations and discuss potential research directions in this field. 
We conclude that integrating LLMs with agent-based simulations offers a powerful toolset for researchers and scientists, allowing for more nuanced, realistic, and comprehensive models of complex systems and human behaviours. 
\end{abstract}


\end{frontmatter}


\input{document}

\bibliographystyle{vancouver} 
\bibliography{sn-bibliography}

\end{document}

%% file: document.tex
\section{Introduction}

Large Language Models (LLMs) have experienced a rapid expansion in adoption across a multitude of research domains and practical applications in recently. 
This swift proliferation is largely attributed to their remarkable ability in understanding, generating, and translating human language with unprecedented accuracy and fluency. 
Industries ranging from healthcare \cite{Clusmann2023,Barrington2023}, where they assist in patient care and medical research, to finance \cite{Wu2023}, for analyzing market trends and automating customer service, have leveraged the capabilities of LLMs to enhance efficiency and innovation. 
Furthermore, in the realm of academia, these models are instrumental in analyzing vast datasets for insights \cite{Taylor2022}, thereby accelerating research outcomes in fields such as social sciences, linguistics, and computer science.
The versatility and evolving sophistication of LLMs have thus cemented their role as a cornerstone technology that is reshaping the landscape of both industry and research, fostering new methodologies and approaches across disciplines.

In social sciences, one area where LLMs can be effectively applied is social simulations \cite{Grossmann2023}.
Social simulations are used to model and analyze the complex interactions within social systems, including factors like individual behaviours, group dynamics, social norms, and institutional structures. 
These simulations aim to understand, predict, or examine hypothetical scenarios within social systems.
In that sense, a technique used for social simulations is agent-based modelling (ABM).
ABM is used to design and simulate the actions and interactions of autonomous agents (individuals, entities, or organizations) to assess their effects on the system as a whole.
ABM can utilize AI techniques to enhance the models' complexity, adaptability, and realism.
So far, ABM has already practically incorporated AI techniques like Machine Learning (ML) \cite{Turgut2023,Dehkordi2023,Chen2021,Vahdati2019}, Reinforcement Learning \cite{Chmura2007,Chen2018} and Inverse Reinforcement Learning \cite{Lee2018} in various social simulation studies.
However, the potential of LLMs in assisting in understanding of complex social systems has yet to be realized.
There are few studies focusing on such incorporation recently.
Even though these studies provide some practical solutions, they lack a well-defined conceptual baseline which is necessary to explore general-purpose architectures and methodologies – eventually extending existing ones – that are effective to seamlessly and systematically integrate social simulations and LLMs.

Furthermore, LLMs can streamline and augment various other aspects of the ABM process, such as literature reviewing, data preparation and interpretation, calibration of parameters, sensitivity analyses, analyses of results.
We have already begun to see some impacts of LLMs on these activities.
But they are scarce and do not provide a clear vision in terms of social simulations.
Drawing upon these observations, in this paper, we outline prospective avenues for research into the augmentation of agent-based simulations through the integration of LLMs. 
We articulate a structured exploration of the symbiotic potential between LLMs and ABM frameworks, aiming to advance the methodological foundations and enhance the analytical capabilities of social simulations.

The remainder of the paper is organized as follows.
We first provide a background on LLMs and ABM for social simulations in Section \ref{sec:Background}.
After, we identify the conceptual baseline by evaluating various methodologies for building multi-agent systems in Section \ref{sec:The-Conceptual-Baseline}.
Then, we provide an overview of some main research directions useful to develop the idea in Section \ref{sec:Research-Directions} and finally in Section \ref{sec:Conclusions} we conclude the paper.


\section{Background}
\label{sec:Background}

\subsection{Large-Language Models (LLMs)}
\label{sec:Large-Language-Models}

A large language model (LLM) can be defined as a function that is finding, considering a series of tokens (such as words, word fragments, punctuation, emojis, etc.), which tokens are most probable to follow next.
As of today, there are numerous LLMs accessible to the public. 
Google's Bard, based on the more efficient and compact PaLM 2 LLM with 340 billion parameters, is accessible for free via a web browser\footnote{\url{https://bard.google.com/chat}, last access on 07/02/2024.} and API, featuring a diverse training dataset \cite{Anil2023}. 
Meta's LLaMA\footnote{\url{https://llama.meta.com}, last access on 07/02/2024.}, aimed at AI research advancement, offers various models with up to 70 billion parameters available to researchers through application \cite{Touvron2023}.
OpenAI's GPT models utilize transformer architecture for dynamic output generation \cite{Vaswani2017}, with various versions accessible differently: GPT-3.5 is free via a web interface\footnote{\url{https://chat.openai.com}, last access on 07/02/2024.}, while GPT-4 requires a subscription, with API usage also being pay-per-use based on tokenization for natural language processing.
LLM APIs are interfaces that allow developers to access the capabilities of these advanced LLMs in their applications. 
APIs can be integrated into applications using any programming language that can make HTTP requests, typically by sending the prompt for the LLM to process along with various parameters that adjusts the LLM's behavior (such as the version of the language model to use and the temperature that controls the randomness of the generated output).

An LLM can be viewed as a non-deterministic simulator with the ability to role-play an endless array of characters. 
Essentially, LLMs can be fine-tuned by exposing them to specific roles so that they can simulate human-like interactions. 
Fine-tuning is achieved by training the model on a curated dataset that embodies the language, knowledge, and nuances of the roles it is expected to perform. 
This process involves adjusting the model's parameters so it better aligns with the patterns, vocabulary, and decision-making processes characteristic of the target roles. 
During fine-tuning, the model learns to prioritize responses that reflect the specific traits, expertise, or persona of the roles in question. 
This is often done by using a smaller, more specialized dataset after the model has been pre-trained on a broad corpus of text, allowing it to adapt its vast general knowledge to more narrowly defined contexts and behaviors.
However, such dataset as well can be relatively large.
 
A well-known architecture for building systems that leverage LLMs on specialized relatively large datasets is called Retrieval Augmented Generation (RAG) \cite{Lewis2021}.
In this architecture, the dataset is broken into chunks (such as a couple of paragraphs or a page) and then those chunks are sent to an LLM and they are turned into a vector.
Each chunk will have a vector (i.e. a series of numbers) which a numeric representation of the essence of that chunk.
After, each time right after a prompt is sent, its vector is calculated as well using the same LLM. 
Then, the closest chunks are found by performing a mathematical comparison between the chunk vectors and the prompt vector.
Finally, these chunks are used as part of the prompt.

\subsection{ABM in Social Simulations}
\label{sec:ABM-in-Social-Simulations}

ABM stands as a pivotal technique in the exploration and understanding of social systems through computational simulations \cite{Macal2016}. 
Social systems, in this context, refer to complex networks of interactions among individuals, institutions, and their environments (which can include factors like individual behaviours, group dynamics, social norms, and institutional structures).
The multifaceted nature of social systems allows individuals to play multiple roles simultaneously (such as a parent, employee, consumer, and citizen in a human society) each with its own set of expectations and norms.

ABM for social systems aims to mimic social processes by simulating the actions and interactions of agents, which represent individuals or entities within these systems, to predict and understand complex phenomena. 
In other words, ABM enhances the study of social systems by offering a bottom-up modeling approach, where the macro-level phenomena of interest emerge from the micro-level interactions of agents. 
This capability is invaluable in social science, where understanding the emergence of complex social phenomena from simple rules of interaction can provide profound insights into the nature of social order, the evolution of norms and institutions, and the dynamics of social change.
Social simulations provide a powerful means to examine hypothetical scenarios, test theories of social behavior and interaction, and explore the potential effects of policy decisions without the ethical and practical constraints of real-world experimentation.
 
However, analyzing the results of ABM to understand social systems presents several challenges, primarily due to the complexity and dynamism inherent in both the models and the systems they seek to represent.
Firstly, ABMs often generate vast amounts of data through the simulation of interactions among numerous agents over time, making it difficult to discern clear patterns or draw straightforward conclusions.
The emergent phenomena, a hallmark of ABM, while valuable for understanding the macro-level outcomes of micro-level behaviors, can complicate analysis as these outcomes are not always predictable or linear.
Secondly, interpreting the results of ABMs in a manner that is meaningful for policy-making or theoretical advancement requires bridging the gap between complex, often technical, model outputs and the conceptual frameworks of social sciences. 
This often demands interdisciplinary collaboration to ensure that the insights generated are both scientifically rigorous and socially relevant. 

Besides, in addition to ABM, social scientists employ various other methods to study social systems. 
Surveys and questionnaires are used to collect large-scale data on individual attitudes, behaviours, and experiences. 
Interviews and ethnography offer in-depth qualitative insights into social phenomena, capturing the nuances of human behaviour and social interactions. 
Case studies provide detailed analysis of specific instances or events, allowing for a deep understanding of complex social processes. 
Experimental and quasi-experimental designs are utilized to establish causal relationships between variables. 
Statistical analysis and computational techniques, including network analysis and data mining, are applied to analyze and model large datasets. 
Lastly, content and discourse analysis are employed to examine communication patterns and the construction of meaning within social contexts. 
Some of these methods are already used in ABM studies.


\section{The Conceptual Baseline for LLM-Augmented Social Simulations}
\label{sec:The-Conceptual-Baseline}

Integrating LLMs into ABMs for social simulations offers a transformational potential for understanding complex social systems.
However, such integration requires a conceptual baseline to connect both domains.
A conceptual baseline is a clear and coherent foundational framework that outlines key concepts, variables, assumptions, and relationships within a given system or model. 
It serves as a reference point for understanding the dynamics and behaviour of the system being studied, and it is subsequently used for model development and analysis.
Such a conceptual baseline can be established using the existing engineering methodologies dedicated to multi-agent systems (MAS). 
A MAS is typically viewed from four primary angles: agent, interaction, environment and organization.

\textit{Agent-oriented methodologies} prioritize individual agents, their autonomy, internal states, and decision-making \cite{Shoham1993,Cuesta2008,Abdalla2021}. They use the Beliefs, Desires, and Intentions (BDI) model for agent cognition and define agents' capabilities and roles. This approach enables detailed modeling of complex, autonomous agents that make decisions based on perceptions and fulfill roles in multi-agent interactions.
%
%
\textit{Interaction-oriented methodologies} focus on agent communication and coordination dynamics, using protocols and messages to define interaction patterns \cite{Chopra2023,Singh2011,Muller1998}. Protocols ensure structured information exchanges, while negotiation and coordination mechanisms support joint decision-making and action synchronization. Social norms and conventions establish a framework for predictable interactions and adherence to shared conduct rules.
%
%
\textit{Environment-oriented methodologies} emphasize the environment's role in agent interactions, focusing on shared resources, stigmergy for indirect coordination, and affordances that dictate agent actions based on their capabilities \cite{Ricci2011,Weyns2006}.
%
%
\textit{Organizational-oriented methodologies} abstract the MAS regarding groups, teams, and broader organizational structures \cite{Abbas2015,Criado2013,Giorgini2006,Dignum2005,Ferber2004,Hubner2002,Roussille2022}. 
These methodologies focus on how agents are arranged and interact within larger entities, defining roles, responsibilities, and relationships through groups and teams. 
Organizational structures, such as hierarchies or networks, dictate the flow of information and control among agents. 
Policies and regulations set out the rules governing behaviour within the organization, guiding agent actions and interactions. 

The organizational-oriented MAS approach is particularly well-suited for modeling social systems and integrating LLMs due to its emphasis on structured interactions and roles within a complex system. 
This approach mirrors the hierarchical and networked nature of social systems, where entities (agents) assume specific roles and responsibilities that are governed by established norms and policies. 
Such an organizational structure enables the clear definition of roles for LLMs, facilitating their integration as agents with specific functions related to language understanding, generation, and processing. This not only enhances the system's ability to mimic human social structures but also leverages LLMs' capabilities in natural language processing to improve communication and coordination among agents. 
By aligning the structural and functional aspects of MAS with the inherent properties of social systems and the strengths of LLMs, the organizational-oriented approach offers a robust framework for capturing the complexity and dynamics of social interactions, making it a superior choice for these applications.
Therefore, we claim that an efficient ABM tool augmented by LLM should support the organization-oriented conceptual baseline.

In that light, we propose defining agents in social simulations as social agents that are role-playing one or several predefined characters \cite{Shanahan2023,Andreas2022,Park2023}.
A social agent's skills are the role playing capacities they have through their interactions with environments, and are used in a community of other role players who also inhabit in these environments.



\section{Research Directions}
\label{sec:Research-Directions}

  
In this section, we explore key research paths that seem pertinent for transforming social simulations with the help of LLMs.

\subsection{Literature Reviews}



The volume of scientific literature is overwhelming \cite{Hutson2020,Wagner2022}, with varying levels of quality across publications. 
There is a necessity for tools that can search, evaluate, and summarize scientific literature both objectively and efficiently.
LLMs can significantly address the challenges of the traditional review process through their advanced capabilities in processing vast amounts of text data efficiently \cite{Wagner2022,Dunn2022,Zhang2023b} and and they are less likely to cherry-pick the literature to support desired hypotheses (i.e. reduced researcher bias) \cite{Muller2022}. 
By automating the initial screening and summarization of literature, LLMs can help researchers navigate the issue of information overload, enabling them to quickly identify relevant studies without compromising the breadth or depth of the review. 
Their ability to analyze and summarize texts in multiple languages can also overcome language barriers, providing access to a wider range of literature. 


\subsection{Modeling Architectures}

As discussed in Section \ref{sec:The-Conceptual-Baseline}, the organization-oriented approach which is defining agents as role playing actors provides a solid grounding for modeling social agents. 
%
%
However, there are several organization-oriented architectures out there and the fit for the LLM-augmented social simulations is yet to be studied. 
%
%
Moreover, research should also be done for the effective design and re-use of roles of social agents.
Effective design of roles is necessary for obtaining effective insights (see Section \ref{sec:Obtaining-Insights}) and effective re-use of roles is necessary for large and repeatable social simulations.
Furthermore, LLMs can be used to generate agent-based models and scenarios using natural language and related qualitative data.

\subsection{Data Preparation}
\label{sec:Data-Preparation}

The data collection for social simulations is time consuming and expensive since it faces several key challenges, including capturing the complexity of social systems, ensuring high-quality and relevant data, addressing ethical and privacy concerns, integrating diverse data sources, and achieving accurate modeling and simulation. 
Additionally, the calibration and validation of models, managing computational constraints, fostering interdisciplinary collaboration, and adapting to dynamic social systems add layers of complexity. 
Finally, ensuring the generalizability and transferability of the models to different contexts or populations is a significant challenge, requiring a careful, methodical approach and often interdisciplinary collaboration.
LLMs have the potential to significantly enhance the data collection process for social simulations by offering solutions to many of the challenges outlined. 
For capturing the complexity of social systems, LLMs can process and analyze large volumes of text data from diverse sources \cite{Zhang2023,Semeler2024}, providing a nuanced understanding of social dynamics that can inform more accurate and comprehensive models. 
Their advanced natural language processing capabilities enable the integration of varied data types, from structured data to unstructured text, facilitating the creation of richer, multidimensional datasets.

\subsection{Dataficiation}
\label{sec:Dataficiation}

Datafication is the transformation of complex social interactions and phenomena into quantifiable data, allowing for real-time tracking and predictive analysis \cite{MayerSchnberger2014}.
LLM-augmented social agents can play a pivotal role in the process of datafication. 
This data can then be analyzed to uncover patterns, trends, and insights about social behaviors and systems.
For instance, social agents interact within simulations, they continuously generate data that can reflect changes over time, including how social systems evolve in response to external pressures or internal dynamics. 
This dynamic aspect of data generation is crucial for studying processes of social change, innovation diffusion, and the emergence of social norms.

\subsection{Obtaining Insights}
\label{sec:Obtaining-Insights}

Social simulations often produce data that are too voluminous and too complex to curate and analyse.
It is shown that using LLMs it is possible to get insights from data \cite{McCloskey2024}.
Hence, insights can be obtained social agents by simply entering into dialogues with them.
If prepared properly, it is possible to simulate a synthetic population of a vast array of social agents representing human experiences and perspectives \cite{Fuchs2023}, which can provide a more precise depiction of human behavior and social dynamics than what is achievable through traditional methods \cite{Grossmann2023}. 
With proper conditioning \cite{Argyle2023}, social agents can role-play characters that have beliefs and intentions and that provide accurate and objective answers to users' questions \cite{Mei2024}. 
They are ideal obtaining insights as they “can rapidly answer hundreds of questions without fatigue” and “need fewer incentives than humans to give reliable responses” \cite{Dillion2023}.
Recent studies show that LLMs can have the ability to make judgements quite well aligned with human judgements \cite{Dillion2023}.
%
%
The information collected from various dialogues can then be organized as quantifiable data that can be statistically analyzed for providing insights into broader social trends and patterns.
Such an approach can then be used in generating hypotheses and validating them in human societies \cite{Argyle2023,Park2023}.

\subsection{Explainability}

LLM-augmented social agents can generate natural language explanations for their actions, decisions, and the underlying mechanics of the simulation \cite{Peterson2021,Gil2017}.
This can make the behavior of these agents more understandable to researchers (from various disciplines, not just those with computational backgrounds), stakeholders, and the general public, translating complex algorithms and decision-making processes into easily digestible explanations.
Moreover, by leveraging the vast knowledge and understanding of social dynamics embedded within LLMs, these augmented agents can provide contextual insights into their behavior. 
For instance, they can explain how certain social norms, historical events, or cultural aspects influence their actions within the simulation, offering a deeper understanding of the modeled social phenomena.


%

\subsection{Platforms and Tools}

Effective LLM-augmented social simulations require sophisticated tool support, tailored to manage the complexities of dynamic social interactions and datafication processes. 
These tools should facilitate the seamless integration of LLMs with simulation platforms, offering features for easy configuration, real-time adjustments, and ethical data handling. Importantly, the design of these tools must be grounded in an organization-oriented conceptual baseline (see Section \ref{sec:The-Conceptual-Baseline}), ensuring they align with the structured dynamics of social systems and support focused, interdisciplinary research. 
This approach enhances the accessibility, accuracy, and ethical compliance of simulations, enabling researchers to deeply explore and understand social phenomena.




\section{Conclusions}
\label{sec:Conclusions}

LLMs offer a transformative framework for the simulation and analysis of social systems.
In this study, we proposed different LLM interventions that span the entire social simulation pipeline.
In addition to reasoning the behavior of involved agents, LLMs can enable more prosperous and intuitive interactions between users and simulated agents.
This can make social simulation more accessible and user-friendly, allowing individuals from various disciplines, including social sciences, healthcare, urban planning, and environmental studies, to utilize them.
Such democratization can enable these professionals to model and analyze complex systems relevant to their fields without needing deep expertise in computer science.
As a result, we can expect an increased collaboration between computer scientists and experts from other fields, leading to more interdisciplinary approaches and innovative solutions to complex real-world problems.

The potential advantages of Large Language Models (LLMs) merit serious consideration. 
However, it is essential for scientists and developers working with LLM-enhanced Agent-Based Modeling (ABM) tools to also entertain the possibility that these tools could, under certain circumstances, hinder rather than advance scientific knowledge. 
This means that while LLMs offer significant epistemic benefits, they also pose epistemic risks if scientists rely on them as partners in producing knowledge \cite{Messeri2024}.
Because, treating LLM-augmented ABM tools as collaborators in scientific research exposes scientists to the risk of falling into illusions of understanding, which is a class of metacognitive error that occurs when individuals have mistaken beliefs about the depth or accuracy of their own comprehension \cite{Rozenblit2002,Rabb2019}.


\section*{Acknowledgements}
The work reported here is part of the URBANE project, which has received funding from the European Union’s Horizon Europe Innovation Action under grant agreement No. 101069782.